\documentclass[conference,onecolumn,letterpaper,final]{IEEEtran}
\usepackage{caption}
\usepackage{subcaption}  
\usepackage{algorithm}
\usepackage{algpseudocode}
\usepackage{algorithm} 
\usepackage{longtable}

\usepackage{titlesec}

\usepackage[T1]{fontenc}
\usepackage{lmodern} 
\usepackage{amssymb,stmaryrd,amsmath,amsfonts,rotating}
\usepackage{amsmath,amsthm,amssymb,cases}
\usepackage{tikz}
\usetikzlibrary{arrows}
\usetikzlibrary{tikzmark}
\usetikzlibrary{matrix}
\usetikzlibrary{positioning}
\usepackage{algpseudocode}
\usepackage{amsmath,cases}
\usepackage{bbm}
\usepackage{amsthm}
\usepackage{booktabs}
\usepackage{color}
\usepackage{amssymb}
\usepackage{bm}
\usepackage{thmtools}

\theoremstyle{definition}
\declaretheoremstyle[style=definition,qed=\openbox,]{ppstyle}

\newtheorem{definition}{Definition}

\usepackage{mathtools}
\mathtoolsset{showonlyrefs}
\declaretheorem[name=Example, style=ppstyle,]{example}

\usepackage{mathtools}
\usepackage{amsmath}

\usepackage{bbm}
\usepackage{mathrsfs}

\makeatletter
\newcommand\RedeclareMathOperator{%
  \@ifstar{\def\rmo@s{m}\rmo@redeclare}{\def\rmo@s{o}\rmo@redeclare}%
}
\newcommand\rmo@redeclare[2]{%
  \begingroup \escapechar\m@ne\xdef\@gtempa{{\string#1}}\endgroup
  \expandafter\@ifundefined\@gtempa
     {\@latex@error{\noexpand#1undefined}\@ehc}%
     \relax
  \expandafter\rmo@declmathop\rmo@s{#1}{#2}}
\newcommand\rmo@declmathop[3]{%
  \DeclareRobustCommand{#2}{\qopname\newmcodes@#1{#3}}%
}
\@onlypreamble\RedeclareMathOperator
\makeatother

\definecolor{grey}{rgb}{0.7, 0.75, 0.71}

\def\dark-red#1{\textcolor[rgb]{0.7,0.0,0.0}{#1}}

\definecolor{amber}{rgb}{1.0, 0.75, 0.0}

\def\...{\dotsc}

\def\intT2{\int_{-T/2}^{T/2}}

\def\sumi1n{\sum_{i=1}^{n}}
\def\sumi1N{\sum_{i=1}^{N}}
\def\sumi0N--{\sum_{i=0}^{N-1}}

\def\=def{\overset{\text{\small def}}{=}}
\newcommand{\defeq}{\overset{\text{\small def}}{=}}

\DeclarePairedDelimiterX{\inp}[2]{\langle}{\rangle}{#1, #2}

\def\Fb{\mathbb{F}}

\def\<{\langle}
\def\>{\rangle}

 \def\mat4#1#2#3#4{
\begin{pmatrix}
 #1&\ccc&#2\\
 \vdots&&\vdots\\
 #3&\ccc&#4
\end{pmatrix}}

\def\0sf{\mathsf{0}}
\def\1sf{\mathsf{1}}

\def\0BS{\boldsymbol{0}}
\def\1BS{\boldsymbol{1}}

\def\0B{\mathbf{0}}
\def\1B{\mathbf{1}}

\def\0H{\hat{0}}
\def\1H{\hat{1}}

\def\+TT{\texttt{+}}
\def\-{\texttt{-}}
\def\+KB{|+\> \<+|}
\def\-KB{|-\> \<-|}

\def\q0{|0\>}

\def\0U{\underline{0}}
\def\1U{\underline{1}}

\def\0UH{\underline{\0H}}
\def\1UH{\underline{\1H}}

\RedeclareMathOperator{\Im}{Im}

\ifCLASSINFOpdf
\else
\fi
\begin{document}
%
\title{Systematic Non-Binary Extension of LDPC-CSS Codes Preserving Orthogonality}
\author{
\IEEEauthorblockN{Kenta Kasai}
\IEEEauthorblockA{
Institute of Science Tokyo\\
Email: kenta@ict.eng.isct.ac.jp}
}
\maketitle

\begin{abstract}
We study finite-field extensions that preserve the same support as the parity-check matrices defining a given binary CSS code.
Here, an LDPC-CSS code refers to a CSS code whose parity-check matrices are orthogonal in the sense that each pair of corresponding rows overlaps in an even (possibly zero) number of positions, typically at most twice in sparse constructions.
Beyond the low-density setting, we further propose a systematic construction method that extends to arbitrary CSS codes, providing feasible finite-field generalizations that maintain both the binary support and the orthogonality condition.
\end{abstract}
\IEEEpeerreviewmaketitle

\section{Introduction}
In this paper, we address the problem of extending a binary CSS code to a finite-field representation while preserving its structural properties.  
Given a pair of binary parity-check matrices whose corresponding rows overlap in an even number of positions, we seek to construct non-binary matrices over a finite field that share the same pattern of nonzero entries and remain orthogonal.  
Although this task can be viewed as a multivariate quadratic feasibility problem over a finite field, it is generally hard to solve due to the nonlinear nature of the constraints and the exponential size of the solution space.

The method originally proposed in~\cite{6017122} transforms a pair of orthogonal binary matrices into non-binary matrices over a finite field while keeping the same support pattern.  
This technique was first developed for quasi-cyclic LDPC codes~\cite{Hagiwara2007} with column weight two, later applied to toric codes~\cite{6284205}, and subsequently generalized to protograph-based constructions~\cite{komoto2024quantumerrorcorrectionnear}.  
The resulting non-binary CSS codes, as demonstrated in recent works~\cite{komoto2024quantumerrorcorrectionnear,kasai2025degeneracy}, achieved decoding performance close to the hashing bound when joint belief propagation was used.

However, these previous approaches were limited to cases where each column contains exactly two ones.  
Extending the construction to more general CSS codes---where column weights may vary across different rows and matrices---remains a challenging problem.  
In such settings, one must simultaneously preserve the binary support and maintain the orthogonality between the two parity-check matrices, while also avoiding short cycles and ensuring sufficient flexibility in assigning non-binary coefficients.

Although orthogonality can be trivially maintained by assigning constant coefficients to each column or by setting all nonzero entries to one, such simple constructions tend to degrade important code properties such as minimum distance and error-floor performance.  
Our goal, therefore, is to characterize the entire space of feasible finite-field extensions so that one can systematically or randomly generate valid non-binary instances with diverse coefficient configurations.

To achieve this, we parameterize the non-binary entries using multiplicative representations over the finite field and translate the orthogonality constraints into a sparse system of linear congruences among integer exponents.  
This system can be efficiently solved either in canonical form using Smith normal decomposition or through a lightweight elimination process involving only integer additions and row swaps.  
Once a feasible set of exponents is obtained, the corresponding non-binary matrices are recovered by standard linear algebra over the finite field.

As a simple baseline, we introduce a separable assignment of coefficients that always guarantees orthogonality for any even-overlap pattern.



\section{Problem Setting}
This section formulates the finite-field extension problem for arbitrary CSS codes. 
Let \((H_C,H_D)\) be a binary orthogonal pair with \(H_C\in\{0,1\}^{M_X\times N}\) and \(H_D\in\{0,1\}^{M_Z\times N}\).
Write the entries of \(H_C\) and \(H_D\) as \(c_{i,j}\) and \(d_{i',j}\), respectively.
They satisfy the binary orthogonality condition
\[
H_C H_D^{\mathsf T}=0,
\]
which is equivalent to
\begin{align}
 \sum_{j=0}^{N-1} c_{i,j}\, d_{i',j}=0
 \quad\text{for all } i=0,\ldots,M_X-1,\; i'=0,\ldots,M_Z-1 .
\label{eq:binary_orthogonality}
\end{align}
All arithmetic in \eqref{eq:binary_orthogonality} is over \(\Fb_2\).

Our goal is to construct matrices \(H_\Gamma=(\gamma_{i,j})\in \Fb_q^{M_X\times N}\) and 
\(H_\Delta=(\delta_{i',j})\in \Fb_q^{M_Z\times N}\) such that
\[
H_\Gamma H_\Delta^{\mathsf T}=0,
\qquad
\mathrm{supp}(H_\Gamma)=\mathrm{supp}(H_C),
\qquad
\mathrm{supp}(H_\Delta)=\mathrm{supp}(H_D).
\]
Equivalently, the support constraints enforce
\[
\gamma_{i,j}=0 \;\;\text{iff}\;\; c_{i,j}=0,
\qquad
\delta_{i',j}=0 \;\;\text{iff}\;\; d_{i',j}=0,
\]
so in particular \(c_{i,j}=1\Rightarrow \gamma_{i,j}\in\Fb_q^{\times}\) and 
\(d_{i',j}=1\Rightarrow \delta_{i',j}\in\Fb_q^{\times}\).

The orthogonality over \(\Fb_q\) is then
\begin{align}
 \sum_{j=0}^{N-1} \gamma_{i,j}\,\delta_{i',j}=0
 \quad\text{for all } i=0,\ldots,M_X-1,\; i'=0,\ldots,M_Z-1 .
\label{eq:orthogonality}
\end{align}
All arithmetic in \eqref{eq:orthogonality} is over \(\Fb_q\).

The unknowns are the entries \(\{\gamma_{i,j}: c_{i,j}=1\}\cup\{\delta_{i',j}: d_{i',j}=1\}\).
The constraints \eqref{eq:orthogonality} yield \(M_XM_Z\) homogeneous quadratic equations in these variables over \(\Fb_q\).
Thus, finding a feasible assignment is an instance of the multivariate quadratic (MQ) feasibility problem,
which is NP-complete~\cite{garey1979computers}.

\section{Finite-Field Extension for LDPC-CSS Codes}
The binary orthogonality condition~\eqref{eq:binary_orthogonality}
implies that every row pair \((i,i')\) has an even number of common nonzero columns.
Let \(S_{i,i'} \subseteq \{0,1,\ldots,N-1\}\) denote
the set of column indices where both \(c_{i,j}=1\) and \(d_{i',j}=1\); that is,
\[
S_{i,i'} \defeq \{\, j \mid c_{i,j}=1 \text{ and } d_{i',j}=1 \,\}.
\]
By construction, \(|S_{i,i'}|\) is always even.

We focus on the canonical case in which each row pair shares either \(0\) or \(2\) columns.
This covers many practical CSS constructions (e.g., quasi-cyclic and protograph-based designs) and
arises frequently in sparse instances, where pairwise overlaps beyond two are unlikely.

Assume \(H_C H_D^{\mathsf T}=0\), fix \(q=2^m\), and let \(\alpha\in\Fb_q^\times\) be a primitive element.
Let \(H_\Gamma=(\gamma_{i,j})\) and \(H_\Delta=(\delta_{i',j})\) preserve the binary support:
each zero in \((H_C,H_D)\) remains zero, and each nonzero entry is replaced by an element of \(\Fb_q^\times\).
We parameterize the nonzero entries multiplicatively as
\[
  \gamma_{i,j}=\alpha^{e_{i,j}}, \qquad \delta_{i',j}=\alpha^{f_{i',j}},
\]
with exponents \(e_{i,j},f_{i',j}\in \mathbb{Z}/(q-1)\mathbb{Z}\).

If a row pair \((i,i')\) shares exactly two columns \(j\) and \(j'\),
the orthogonality condition \(\sum_j \gamma_{i,j}\delta_{i',j}=0\) over \(\Fb_q\) reduces to
\[
  \gamma_{i,j}\delta_{i',j}+\gamma_{i,j'}\delta_{i',j'}=0.
\]
Since \(\mathrm{char}(\Fb_q)=2\), we have \(x+y=0 \iff x=y\).
Hence
\[
  \gamma_{i,j}\delta_{i',j}=\gamma_{i,j'}\delta_{i',j'}
  \;\Longleftrightarrow\;
  \alpha^{\,e_{i,j}-e_{i,j'}+f_{i',j}-f_{i',j'}}=1,
\]
and therefore
\[
  e_{i,j}-e_{i,j'}+f_{i',j}-f_{i',j'}\equiv 0 \pmod{q-1}.
\]
Collecting these relations over all row pairs \((i,i')\) with \(|S_{i,i'}|=2\)
produces a sparse system of linear congruences:
\begin{equation}
  A v \equiv 0 \pmod{q-1},
  \label{eq:exponent_congruences}
\end{equation}
where \(v\) stacks the unknown exponents \(\{e_{i,j}:c_{i,j}=1\}\cup\{f_{i',j}:d_{i',j}=1\}\).
The coefficient matrix \(A\in\{0,\pm1\}^{r\times n}\) has one row for each overlapping pair \((i,i';j,j')\)
and one column for each nonzero entry of \(H_C\) or \(H_D\).
Equation~\eqref{eq:exponent_congruences} compactly captures all additive consistency constraints among the exponents.

\subsection{Solving the Linear Congruence System}
For a modulus \(n=q-1>0\) and an integer matrix \(A\),
the homogeneous system \(A v \equiv 0\pmod{n}\)
can be analyzed via the Smith normal form (SNF) of \(A\) over \(\mathbb{Z}\)~\cite{smith1861xv},
which provides a canonical decomposition of the solution space.

When \(q-1\) is prime (that is, when \(2^m-1\) is a Mersenne prime),
the ring \(\mathbb{Z}/(q-1)\mathbb{Z}\) forms a finite field isomorphic to \(\Fb_{q-1}\).
In this case, the system \(A v \equiv 0\) is an ordinary linear equation over \(\Fb_{q-1}\),
and its nullspace can be computed by standard Gaussian elimination.

Even when \(q-1\) is composite, it is often possible to solve \(A v \equiv 0\pmod{q-1}\)
efficiently without computing the SNF.
This is because all nonzero coefficients of \(A\) are in \(\{+1,-1\}\),
so that elimination can proceed using only row swaps and
addition or subtraction of rows modulo \(q-1\), without any division.
In practice, this yields a fraction-free elimination process that produces
a row-reduced form with unit pivots (each pivot being \(1\) or \(-1\)),
after which the nullspace can be read off by back-substitution.

The elimination proceeds as follows:
(1) at each step, choose a pivot column containing an entry \(\pm1\),
and swap that row into the pivot position;
(2) for every other row having a nonzero in the same column,
add or subtract the pivot row to zero out that entry;
(3) once an echelon pattern is obtained, clear the entries above each pivot
using the same \(\pm\) row additions;
and (4) identify the free variables corresponding to columns without pivots.


In our experiments, every sparse LDPC-CSS instance satisfied this condition,
and all systems were successfully reduced to a row-reduced form
using only row swaps and \(\pm\) additions, without any modular division.

Although a formal guarantee of solvability has not been established, 
we also experimented with a heuristic iterative approach that exploits the sparsity of \(A\).  
In this method, given randomly initialized exponents \(e_j\) and \(f_j\), 
we successively update individual variables in an iterative fashion so that the congruence relations are gradually satisfied.  
Within the scope of our experiments, this procedure consistently produced feasible solutions.

\begin{example}
We briefly recall the hypergraph-product (HGP) construction~\cite{TillichZemor2014}.
Let $H_1\in\{0,1\}^{r_1\times n_1}$ and $H_2\in\{0,1\}^{r_2\times n_2}$.
Define the CSS pair
\begin{align}
 H_X & =\left(\begin{array}{ll}H_1 \otimes I_{n_2} & I_{r_1} \otimes H_2^T\end{array}\right) \\ 
H_Z & =\left(\begin{array}{ll}I_{n_1} \otimes H_2 & H_1^T \otimes I_{r_2}\end{array}\right)
\end{align}
with overall block length $n=n_1 n_2+r_1 r_2$.
One checks $H_X H_Z^{\mathsf T}=0$ by the mixed-product property of the Kronecker product.
This product code has non-vanishing rate and minimum distance scaling on the order of $\sqrt{n}$ when $H_1,H_2$ are good classical LDPCs~\cite{TillichZemor2014}.

Binary inputs \(H_1, H_2\) (column-weight unconstrained):
\[
H_1 =
\begin{bmatrix}
1 & 1 & 1\\
0 & 1 & 0
\end{bmatrix},
\quad
H_2 =
\begin{bmatrix}
1 & 0 & 0\\
1 & 1 & 1
\end{bmatrix}.
\]

Binary HGP matrices \((H_X, H_Z)\):
\[
H_X =
\left(
\begin{array}{ccc|ccc|ccc||cc|cc}
1&0&0&1&0&0&1&0&0&1&1&0&0\\
0&1&0&0&1&0&0&1&0&0&1&0&0\\
0&0&1&0&0&1&0&0&1&0&1&0&0\\
\hline
0&0&0&1&0&0&0&0&0&0&0&1&1\\
0&0&0&0&1&0&0&0&0&0&0&0&1\\
0&0&0&0&0&1&0&0&0&0&0&0&1
\end{array}
\right),
\quad
H_Z =
\left(\begin{array}{ccc|ccc||ccc|cc|cc}
1&0&0&0&0&0&0&0&0&1&0&0&0\\
1&1&1&0&0&0&0&0&0&0&1&0&0\\
\hline
0&0&0&1&0&0&0&0&0&1&0&1&0\\
0&0&0&1&1&1&0&0&0&0&1&0&1\\
\hline
0&0&0&0&0&0&1&0&0&1&0&0&0\\
0&0&0&0&0&0&1&1&1&0&1&0&0
\end{array}\right).
\]
By construction, \(H_X H_Z^{\mathsf T} \equiv 0~(\bmod~2)\).  
Several columns have weights ranging from one to three.  

Fix \(q = 256\) and a primitive element \(\alpha \in \mathbb{F}_{256}^\times\).  
For each nonzero entry of \(H_\Gamma\) and \(H_\Delta\), write \(\gamma_{i,j} = \alpha^{e_{i,j}}\) and \(\delta_{i',j} = \alpha^{f_{i',j}}\), respectively.  
Then, for every row pair \((i, i')\) with \(S_{i,i'} = \{j, j'\}\), the orthogonality condition implies  
\[
e_{i,j} - e_{i,j'} + f_{i',j} - f_{i',j'} \equiv 0 \pmod{255}.
\]

Let \(q=256\) so that \(q-1=255\).
Solving the sparse congruence system~\eqref{eq:exponent_congruences} yields exponent pairs
\((e_{i,j}), (f_{i',j})\) modulo \(255\), which imply the following linear relations (a representative subset):
\begin{align*}
e_{0,0} - e_{0,9} + f_{0,0} - f_{0,9}\equiv 0\\
e_{0,0} - e_{0,10} + f_{1,0} - f_{1,10}\equiv 0\\
e_{0,3} - e_{0,9} + f_{2,3} - f_{2,9}\equiv 0\\
e_{0,3} - e_{0,10} + f_{3,3} - f_{3,10}\equiv 0\\
e_{0,6} - e_{0,9} + f_{4,6} - f_{4,9}\equiv 0\\
e_{0,6} - e_{0,10} + f_{5,6} - f_{5,10}\equiv 0\\
e_{1,1} - e_{1,10} + f_{1,1} - f_{1,10}\equiv 0\\
e_{1,4} - e_{1,10} + f_{3,4} - f_{3,10}\equiv 0\\
e_{1,7} - e_{1,10} + f_{5,7} - f_{5,10}\equiv 0\\
e_{2,2} - e_{2,10} + f_{1,2} - f_{1,10}\equiv 0\\
e_{2,5} - e_{2,10} + f_{3,5} - f_{3,10}\equiv 0\\
e_{2,8} - e_{2,10} + f_{5,8} - f_{5,10}\equiv 0\\
e_{3,3} - e_{3,11} + f_{2,3} - f_{2,11}\equiv 0\\
e_{3,3} - e_{3,12} + f_{3,3} - f_{3,12}\equiv 0\\
e_{4,4} - e_{4,12} + f_{3,4} - f_{3,12}\equiv 0\\
e_{5,5} - e_{5,12} + f_{3,5} - f_{3,12}\equiv 0
\end{align*}

An explicit feasible pair \((H_\Gamma,H_\Delta)\) (entries shown in hexadecimal for \(\Fb_{256}\); \texttt{00} denotes \(0\)) is
\begin{equation}
\begin{aligned}
 H_\Gamma &=
\left[ 
\begin{array}{c@{\hspace{0.5mm}}c@{\hspace{0.5mm}}c|@{\hspace{0.5mm}}c@{\hspace{0.5mm}}c@{\hspace{0.5mm}}c|@{\hspace{0.5mm}}c@{\hspace{0.5mm}}c@{\hspace{0.5mm}}c||@{\hspace{0.5mm}}c@{\hspace{0.5mm}}c|@{\hspace{0.5mm}}c@{\hspace{0.5mm}}c@{\hspace{0.5mm}}}
\mathtt{E9}&\mathtt{00}&\mathtt{00}&\mathtt{F9}&\mathtt{00}&\mathtt{00}&\mathtt{A0}&\mathtt{00}&\mathtt{00}&\mathtt{51}&\mathtt{BE}&\mathtt{00}&\mathtt{00}\\
\mathtt{00}&\mathtt{6A}&\mathtt{00}&\mathtt{00}&\mathtt{0B}&\mathtt{00}&\mathtt{00}&\mathtt{19}&\mathtt{00}&\mathtt{00}&\mathtt{C9}&\mathtt{00}&\mathtt{00}\\
\mathtt{00}&\mathtt{00}&\mathtt{6D}&\mathtt{00}&\mathtt{00}&\mathtt{01}&\mathtt{00}&\mathtt{00}&\mathtt{11}&\mathtt{00}&\mathtt{DA}&\mathtt{00}&\mathtt{00}\\
\hline
\mathtt{00}&\mathtt{00}&\mathtt{00}&\mathtt{EF}&\mathtt{00}&\mathtt{00}&\mathtt{00}&\mathtt{00}&\mathtt{00}&\mathtt{00}&\mathtt{00}&\mathtt{54}&\mathtt{2C}\\
\mathtt{00}&\mathtt{00}&\mathtt{00}&\mathtt{00}&\mathtt{B0}&\mathtt{00}&\mathtt{00}&\mathtt{00}&\mathtt{00}&\mathtt{00}&\mathtt{00}&\mathtt{00}&\mathtt{E6}\\
\mathtt{00}&\mathtt{00}&\mathtt{00}&\mathtt{00}&\mathtt{00}&\mathtt{B7}&\mathtt{00}&\mathtt{00}&\mathtt{00}&\mathtt{00}&\mathtt{00}&\mathtt{00}&\mathtt{09}
\end{array}
\right],
 H_\Delta =
\left[ 
\begin{array}{c@{\hspace{0.5mm}}c@{\hspace{0.5mm}}c|@{\hspace{0.5mm}}c@{\hspace{0.5mm}}c@{\hspace{0.5mm}}c|@{\hspace{0.5mm}}c@{\hspace{0.5mm}}c@{\hspace{0.5mm}}c|@{\hspace{0.5mm}}c@{\hspace{0.5mm}}c|@{\hspace{0.5mm}}c@{\hspace{0.5mm}}c@{\hspace{0.5mm}}}
\mathtt{E2}&\mathtt{00}&\mathtt{00}&\mathtt{00}&\mathtt{00}&\mathtt{00}&\mathtt{00}&\mathtt{00}&\mathtt{00}&\mathtt{7B}&\mathtt{00}&\mathtt{00}&\mathtt{00}\\
\mathtt{A5}&\mathtt{30}&\mathtt{3E}&\mathtt{00}&\mathtt{00}&\mathtt{00}&\mathtt{00}&\mathtt{00}&\mathtt{00}&\mathtt{00}&\mathtt{D0}&\mathtt{00}&\mathtt{00}\\
\hline
\mathtt{00}&\mathtt{00}&\mathtt{00}&\mathtt{7A}&\mathtt{00}&\mathtt{00}&\mathtt{00}&\mathtt{00}&\mathtt{00}&\mathtt{23}&\mathtt{00}&\mathtt{16}&\mathtt{00}\\
\mathtt{00}&\mathtt{00}&\mathtt{00}&\mathtt{C6}&\mathtt{C0}&\mathtt{DB}&\mathtt{00}&\mathtt{00}&\mathtt{00}&\mathtt{00}&\mathtt{02}&\mathtt{00}&\mathtt{8A}\\
\hline
\mathtt{00}&\mathtt{00}&\mathtt{00}&\mathtt{00}&\mathtt{00}&\mathtt{00}&\mathtt{63}&\mathtt{00}&\mathtt{00}&\mathtt{B2}&\mathtt{00}&\mathtt{00}&\mathtt{00}\\
\mathtt{00}&\mathtt{00}&\mathtt{00}&\mathtt{00}&\mathtt{00}&\mathtt{00}&\mathtt{0F}&\mathtt{A1}&\mathtt{BA}&\mathtt{00}&\mathtt{F0}&\mathtt{00}&\mathtt{00}
\end{array}
\right].
\end{aligned}
\end{equation}

One verifies that \(H_\Gamma H_\Delta^{\mathsf T}=0\) over \(\Fb_{256}\) and that all listed congruences are satisfied.
Here, hexadecimal \texttt{00} denotes \(0\),
while \(\mathtt{01}=\alpha^{0}\), \(\mathtt{02}=\alpha^{1}\), and \(\mathtt{10}=\alpha^{15}\)
according to our chosen primitive element \(\alpha\in\Fb_{256}^\times\).

\end{example}

\section{Canonical Feasible Assignment for Any Even Overlap}
\label{subsec:canonical_assignment}
In this section, we assume that the finite field has characteristic two.  
We present a constructive example of a finite-field extension that 
always satisfies the orthogonality condition for any CSS pair whose corresponding rows 
overlap in an even number of positions.  
This baseline serves as a \emph{feasibility witness}, showing that for any even-overlap pattern, 
it is indeed possible to assign finite-field coefficients that guarantee orthogonality 
without solving the full system of multivariate equations.  
Although this construction does not improve the code distance, it provides 
a useful canonical form that clarifies the algebraic structure underlying feasible non-binary assignments.

We first present a baseline \emph{separable} assignment that witnesses feasibility for
any CSS pair whose row–row overlaps are even.
\begin{definition}[Canonical separable assignment (CSA)]
Define
\begin{align}
  \gamma_{i,j} = \alpha^{A_i + C_j},
  \qquad
  \delta_{i',j} = \alpha^{B_{i'} - C_j},
\label{eq:canonical_assignment}
\end{align}
where \(A_i,B_{i'},C_j \in \mathbb{Z}_{q-1}\) and \(\alpha\) is a primitive element of \(\Fb_q\).
\end{definition}

For any \((i,i')\), the product \(\gamma_{i,j}\delta_{i',j}=\alpha^{A_i+B_{i'}}\) is independent of \(j\).
Let \(S_{i,i'}\) be the set of shared columns between row \(i\) of \(H_C\) and row \(i'\) of \(H_D\).
Then, over \(\Fb_q\),
\[
 \sum_{j\in S_{i,i'}} \gamma_{i,j}\delta_{i',j}
 = |S_{i,i'}|\;\alpha^{A_i+B_{i'}} = 0,
\]
because \(|S_{i,i'}|\) is even and \(\mathrm{char}(\Fb_q)=2\).
Hence \(H_\Gamma H_\Delta^{\mathsf T}=0\) holds for any even overlap size
\(|S_{i,i'}|\in\{0,2,4,\ldots\}\), subsuming in particular the common \(0/2\)-overlap case.

While \eqref{eq:canonical_assignment} provides a simple and verifiable construction, 
we present an example where this construction leads to an undesirable result.  

We define
\(C_X = \ker(H_X),
C_Z = \ker(H_Z),
C_\Gamma = \ker(H_\Gamma),
C_\Delta = \ker(H_\Delta). 
\)
It was shown in~\cite{kasai2025degeneracy} that extending to higher-order finite fields can effectively eliminate small-weight logical operators.  
Even if \(\bm{x} \in C_X \setminus C_Z^\perp\), 
the nonzero entries of \(H_\Gamma\) were arranged so that 
there exists no vector \(\bm{\xi}\) having exactly the same support as \(\bm{x}\) and satisfying 
\(\bm{\xi} \in C_\Gamma \setminus C_\Delta^\perp\).  
By doing so, we were able to increase the minimum distance of the code.

However, this mechanism does not apply when the coefficients are assigned according to the CSA, 
since its separable structure preserves all binary codewords up to scalar multiples.  
In this case, for any \(\bm{x} \in C_X\), 
there always exists a vector \(\bm{\xi}\) with the same support satisfying 
\(\bm{\xi} \in C_\Gamma\).  
Specifically, one can take
\[
\xi_j = \alpha^{-C_j} x_j.
\]
Using the same argument as in Section~\ref{subsec:canonical_assignment}, we obtain
\[
(H_\Gamma \bm{\xi}^{\mathsf T})_i
 = \sum_{j: c_{i,j}=1} \alpha^{A_i + C_j}\,\xi_j
 = \sum_{j: c_{i,j}=1} \alpha^{A_i + C_j}\,\alpha^{-C_j} x_j
 = \alpha^{A_i} \sum_{j: c_{i,j}=1} x_j
 = 0,
\]
since \(\bm{x} \in C_X\).
Therefore, \(\bm{\xi}\in C_\Gamma\).  
Hence, to eliminate such short logical operators, one must instead employ the approach based on solving the exponent congruence equations described in the previous section.
\section{Conclusion}

We have presented a systematic method for extending binary CSS codes to non-binary finite-field representations
while preserving both the original support pattern and the orthogonality condition.  
By formulating the orthogonality constraint as a sparse system of linear congruences among integer exponents, 
we showed that feasible non-binary assignments can be efficiently obtained.

As a baseline, we introduced the CSA, which guarantees feasibility for any CSS pair with even row–row overlaps.  
While the CSA provides a simple and verifiable construction, it preserves all binary codewords and thus cannot remove short logical operators.  
To overcome this limitation, we emphasized the importance of solving the exponent congruence equations and applying local rank conditions on small supports, which enable the elimination of low-weight logical operators in the non-binary extension.

The proposed finite-field framework clarifies the algebraic structure underlying feasible non-binary CSS codes and opens up new possibilities for designing high-performance quantum LDPC and qudit codes.  
Future work includes applying this approach to HGP codes as well as to higher-degree quantum APM-LDPC and QC-LDPC codes~\cite{komoto2025sharperrorratetransitionsquantum}, and evaluating their joint belief-propagation decoding performance.

\bibliographystyle{IEEEtran}
\bibliography{IEEEabrv,../../literature/00kasai} 
\end{document}